\input{aipcheck.tex}
\documentclass[final]{aipproc}

\layoutstyle{6x9}

\newcommand\doingARLO[2][]{%
  \ifx\mmref\undefined #1\else #2\fi
}
\newcommand{\Dslash}{{D\hskip -0.23cm\slash}}

\begin{document}

\title
      {A review on axions and the strong CP problem}


\author{Jihn E. Kim}{
  address={Department of Physics and Astronomy, Seoul National University, Seoul 151-747, Korea},
  email={jekim@ctp.snu.ac.kr},
  thanks={This work is supported in part by the Korea Research Foundation, Grant No. KRF-2005-084-C00001.}
}

\iftrue

\fi

\copyrightyear  {2001}

\begin{abstract}
We review the recent developments on axion physics. Some new comments on the strong CP problem, the axion mass, and the simple energy loss mechanism of white dwarfs and related issues are included.
\end{abstract}

\date{\today}

\maketitle

\section{Introduction}

The most awaited information in the universe now is
what is the DM of the universe. One  plausible candidate
is the weakly interacting massive particle(WIMP) and the other attractive candidate is a very light axion. We try to discuss these aspects from
scarce experimental hints and comment a viable model
in a SUSY framework. Axion is a Goldstone boson arising when the PQ
global symmetry is spontaneously broken. The simple
form dictates that its interaction is only through the gluon
anomaly term $G\tilde G$. The axion models have the spontaneous symmetry breaking scale $F$ and the axion decay constant $F_a$ which are related by
$F=N_{\rm DW} F_a$.

The WIMP was first discussed by Lee and Weinberg \cite{LeeWein77} where
a heavy neutrino was considered, which was the beginning of the usage
``weak" in WIMP. The LSP interaction is ``weak" if interaction mediators the SUSY particles are in the 100 GeV range as $W$ boson. That is the reason we talk about WIMP in TeV SUSY. At present, WIMP almost means the LSP.

\begin{figure}
\caption{In (a), a rough sketch of masses and cross
sections modified from \cite{Rosz04} are shown. In (b), the axion among these is shown in the $F_a$ versus the initial misalignment angle plane \cite{BaeFa08}. }\label{SUSYfig02}
\hskip -1cm
\includegraphics[height=.3\textheight]{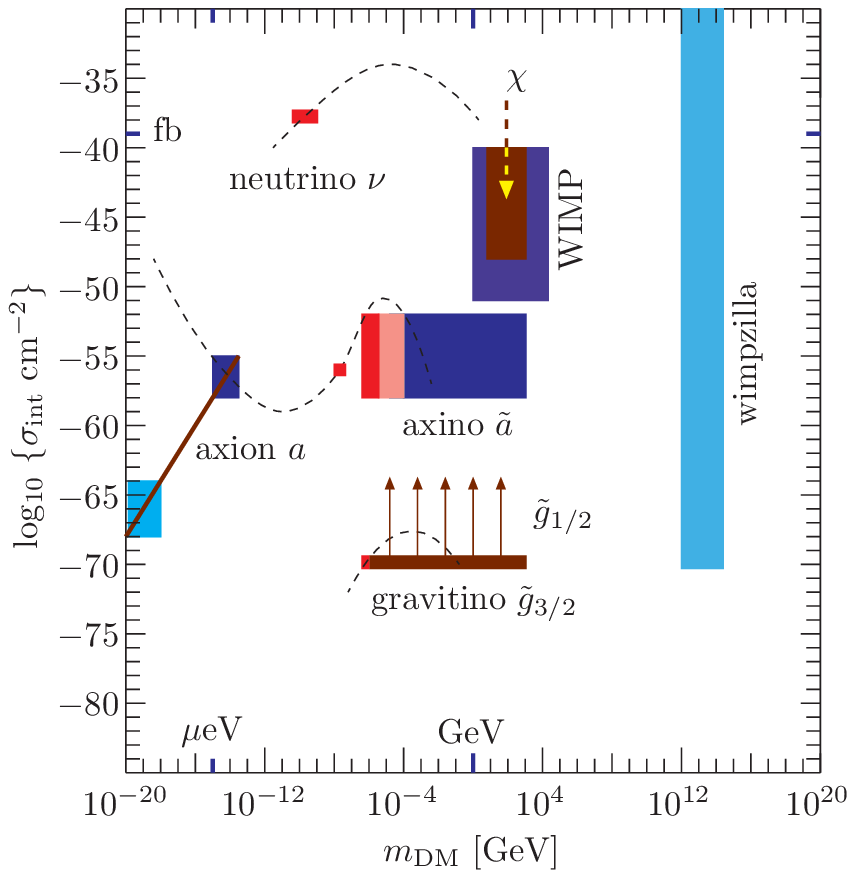}\hskip 1cm
\includegraphics[height=.3\textheight]{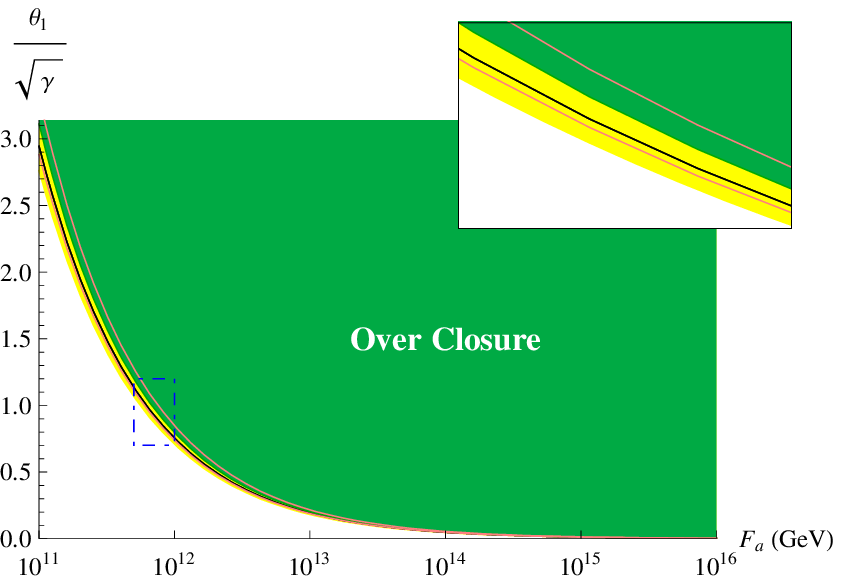}
\hskip -9cm {\small (a)}\hskip 7cm{\small (b)}
\end{figure}

On the other hand, if there exists a coherently oscillating boson field in the universe, this bosonic collective motion is always equivalent to CDM \cite{PWW83}. The dashed lines in Fig. \ref{SUSYfig02}(a) represent the shapes of CDMs of the WIMP and the bosonic collective motion.

Regarding the coherently oscillating bosonic field, the mostly discussed candidate is the very long lived axion \cite{KSVZ,DFSZ}, suggested for the solution of the strong CP problem. The strong CP problem is that QCD allows a CP violating flavor singlet interaction  which however is contradicted by the stringent bound \cite{EDMn} on the electric dipole moment of neutron(EDMn).
The existence of instanton solution in nonabelian gauge
theories needs  $\theta$ vacuum \cite{Theta76}. It introduces an effective interaction, the so-called  $\theta$ term,
\begin{equation}
\frac{\theta}{32\pi^2}G^a_{\mu\nu}\tilde G^{a\mu\nu}
\end{equation}
where $\theta$ is the final value taking into account the electroweak CP violation and $G^a_{\mu\nu}$ is the field strength of gluon.  For QCD to become a correct theory, this CP violating term must be sufficiently suppressed.
Recently, taking the researches of last 20 years, the strong CP problem is
reviewed in Ref. \cite{KimRMP}. Our interest here is the bound on the axion decay constant
\begin{equation}
10^9~ {\rm GeV} \le F_a \le 10^{12}~ {\rm GeV}.
\end{equation}
But the bound is in fact a two-dimensional region as shown in Fig. \ref{SUSYfig02}(b) \cite{BaeFa08}, taking into account the anharmonic term carefully and the new mass bounds on light quark masses. Related studies can be found in \cite{Turner86}.
Fig. \ref{SUSYfig02}(b) is the basis of using the anthropic
argument for a large $F_a$ \cite{Tegmark06}.

\section{Strong CP problem and solutions}
The so-called $\theta$ term resulting from the instanton background led to a sizable EDMn. But the observed EDMn is very tightly bounded, $|d_n/e|<2.9\times 10^{-26}$ cm \cite{EDMn}. In Ref. \cite{KimRMP}, a resulting $\theta$ bound has been given from an effective Lagrangian for the EDMn together with the magnetic dipole moment of neutron(MDMn). The fact that the mass term and the MDMn term have the same chiral transformation property has been used.
The new bound is $|\theta|<0.7\times 10^{-11}$ which is a factor smaller than earlier bounds \cite{Crewther79}. The strong CP problem is given here, ``Why is $\theta$ term so small?" The strong CP problem has been solved in three different categories: (1) Calculable $\theta$, (2) Massless up quark possibility, and
(3) Axion solution.
\begin{itemize}
\item Calculable $\theta$: The calculable solutions were very popular in 1978 \cite{Calculable}. But now  only the Nelson-Barr model \cite{NelsonBarr84} has been remaining, which introduces vectorlike heavy quarks at high energy scale. This model produces the Kobayashi-Maskawa type weak CP violation at the low energy standard model(SM). Still, at one loop the appearance of $\theta$ must be forbidden up to one loop.
\item Massless up quark: Suppose that we chiral-transform a quark $q$ by $e^{i\gamma_5\alpha}q$,
    \begin{equation}\label{eq:massless}
    \int \left(-m\bar qq+\frac{\theta}{32\pi^2}G\tilde G\right) \to \int
     \left(-m\bar q e^{2i\gamma_5\alpha}q+\frac{\theta-2\alpha}{32\pi^2}G\tilde G\right)
    \end{equation}
If $m=0$, it is equivalent to changing $\theta\to\theta-2\alpha$. Thus, there exists a shift symmetry of $\theta\to\theta-2\alpha$. Then, $\theta$ is not physical, and there is no strong CP problem. The physical problem is, ``Is massless up quark phenomenologically viable?" \cite{Kapmassless}. The recent compilation of the light quark masses gives, $m_u=2.5\pm 1$ and $m_d=5.1\pm 1.5$ in units of MeV \cite{Manohar08}, even not considering the lattice calculation of $m_u\ne 0$ \cite{lattice03}. This is convincing enough
that $m_u=0$ is not a strong CP solution.
\item  The axion solution is given below.
\end{itemize}

\section{Axions}\label{sec:Axions}

Historically, Peccei and Quinn (PQ) introduced the so-called PQ symmetry to mimic the symmetry of a massless quark of (\ref{eq:massless}), $\theta\to \theta-2\alpha$, in the full electroweak theory \cite{PQ77}. The PQ symmetry includes the transformations of two Higgs doublets $H_u$ and $H_d$, coupling to up-type and down-type quarks, respectively,
\begin{equation}
q\to e^{i\gamma_5\alpha}q,\ \{H_u,H_d\}\to e^{i\beta} \{H_u,H_d\}
\end{equation}
for a symbolic Lagrangian, ${\cal L}=(\overline{q}_L H_uu_R+\overline{q}_L H_dd_R +{\rm h.c.})-V(H_u,H_d)+\theta\{F\tilde F\}$, which changes to  ${\cal L}=(e^{i(\beta-\alpha)}\overline{q}_L H_uu_R+ e^{i(\beta-\alpha)}\overline{q}_L H_dd_R +{\rm h.c.})-V(H_u,H_d) +(\theta-2\alpha)\{F\tilde F\}$. Therefore, for $\beta=\alpha$ PQ achieved almost the same thing as the massless quark case.
The Lagrangian is invariant under changing  $\theta\to \theta-2\alpha$; thus it seems that $\theta$ is not physical, since it is a phase of the PQ transformation. But, $\theta$ is physical. At the Lagrangian level, there seems to be no strong CP problem. But the VEVs $\langle H_u\rangle$ and $\langle H_d\rangle$ break the PQ global symmetry and there results a Goldstone boson, axion $a$ \cite{PQWW}. Since $\theta$ is made a field, the original  $\cos\theta$ dependence becomes the potential of the axion $a$. If its potential is of the $-\cos\theta$ form, always  $\theta=a/F_a$ can be chosen at 0.  So the PQ solution of the strong CP problem is that the vacuum chooses \cite{PQ77,VW84}
\begin{equation}
\theta=0.
\end{equation}

A historical note is that the above Peccei-Quinn-Weinber-Wilczek(PQWW) axion is ruled out early in one year \cite{Peccei78}. Two years later after many tries of calculable solutions, the PQ symmetry came back incorporating heavy  quarks $Q$, using a singlet Higgs field \cite{KSVZ},
\begin{equation}
{\cal L}=(\overline{Q}_L SQ_R+{\rm h.c.})-V(S,H_u,H_d)+\theta\{F\tilde F\}
\end{equation}
where the Higgs doublets are neutral under the PQ transformation. If they are not neutral, then it is not necessary to introduce heavy quarks but a nontrivial PQ transformation for  the Higgs doublets \cite{DFSZ}. In any case, the axion is the phase of the SM singlet $S$ if the VEV of $S$ is much above the electroweak scale. The couplings of $S$ determine the axion interactions. Because it is a Goldstone boson, the couplings are of the derivative form except the anomaly term.

\begin{figure}
\caption{Contraction of quarks of the 't Hooft determinental interaction. }\label{SUSYfig03}
\includegraphics[height=.25\textheight]{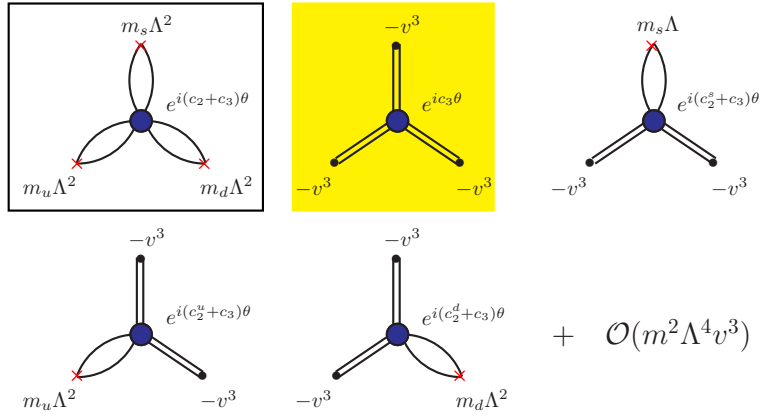}
\end{figure}

In most studies on axions, a specific example is chosen. Here, we consider a general description in an effective theory framework just above the QCD scale. Heavy fermions carrying color charges are special. All heavy fields, including $Q$ and real $S$ are integrated out. So consider the following Lagrangian,
\begin{eqnarray}
{\cal L}_{\theta}&=&\frac12 f_{S}^2\partial^\mu \theta\partial_\mu\theta-\frac1{4g_c^2} G_{\mu\nu}^a G^{a\hskip 0.02cm \mu\nu}+(\bar q_L i\Dslash q_L+\bar q_R i\Dslash q_R)
+ c_1(\partial_\mu \theta)\bar q\gamma^\mu\gamma_5 q\nonumber\\
&&-\left(\bar q_L~ m~q_R e^{ic_2\theta}+{\rm h.c.}\right)
+c_{3} \frac{\theta}{32\pi^2}G^{a}_{\mu\nu}\tilde G^{a\hskip 0.02cm\mu\nu}\ ({\rm or}\ {\cal L}_{\rm det} )\label{Axionint}\\
&&+c_{\theta\gamma\gamma}\frac{\theta}{32\pi^2}F
^{i}_{\rm em,\mu\nu}\tilde F_{\rm em}^{i\hskip 0.02cm\mu\nu}+{\cal L}_{\rm leptons,\theta}\nonumber
\end{eqnarray}
where $\theta=a/f_S$ with the axion decay constant $f_S$ up to the domain wall number ($f_S=N_{DW}F_a$), $q$ is the fermion matrix composed of SU(3)$_c$ charge carrying fields. There are three classes of couplings: $c_1,c_2$ and $c_3$. But
the axion mass depends only on the combination of
$c_2+c_3$ \cite{KimRMP}. The term ${\cal L}_{\rm det}$ is the 't Hooft determinental interaction which is the basis of the solution of the U(1) problem. The determinental interaction is shown in Fig. \ref{SUSYfig03} where the U(1) solution is given pictorially in the second diagram. If the story
ends here, the axion is exactly massless. But, as shown in the other diagrams  in Fig. \ref{SUSYfig03}, there are additional contributions which make axion massive. A $3\times 3$ mass matrix of $\pi^0,\eta',$ and axion can be diagonalized to give \cite{KimRMP}
\begin{equation}
m^2_{\pi^0}=\frac{m_+v^3+2\mu\Lambda_{\rm inst}^3}{f_\pi^2},\ m^2_{\eta'}=\frac{4\Lambda_{\eta'}^4+m_+v^3+2\mu\Lambda_{\rm inst}^3}{f_{\eta'}^2},\ m^2_{a}=\frac{Z}{(1+Z)^2}\frac{f_\pi^2m_{\pi^0}^2(1+\Delta)}{F_a^2}
\end{equation}
where we used the parameters defined in \cite{KimRMP}. The U(1) problem is solved by the first term of $m^2_{\eta'}$. In the axion mass, $\Delta$ represents the instanton contribution and $F_a=F/|c_2+c_3|$ in terms of the singlet VEV scale $F$. Numerically, the axion mass in units of eV is $\simeq 0.6\times 10^{7}{~ \rm GeV}/F_a$.
The essence of the axion solution is that $\langle a\rangle$ seeks $\theta=0$
whatever happened before. In this sense it is a cosmological solution. The height of the potential is guessed as the QCD scale $\Lambda_{\rm QCD}^4$.

\subsection{Axion couplings}

Above the electroweak scale, we integrate out heavy fields. If colored quarks are integrated out, its effect is appearing as the coefficient of the gluon anomaly. If only bosons are integrated out as in the DFSZ model, there is no such anomaly term. Thus, we have
\begin{eqnarray}
&{\rm KSVZ}:& c_1=0,\ c_2=0,\ c_3={\rm nonzero}\nonumber\\
&{\rm DFSZ}:& c_1=0,\ c_2={\rm nonzero},\ c_3=0\\
&{\rm PQWW}:& {\rm similar~ to~ DFSZ}\nonumber
\end{eqnarray}
The axion-hadron couplings are important for the study of supernovae:
The chiral symmetry breaking is properly taken into account,
using the reparametrization invariance so that $c_3'=0$ so that the axion-quark couplings are
\begin{eqnarray}
{\rm KSVZ}:& \bar c_1^{u,d}=\frac12 \bar c_2^{u,d},\\
{\rm DFSZ}:&  \bar c_1^{u}=-\frac{v_d^2}{2(v_u^2+v_d^2)}+\frac12 \bar c_2^{u,d},\  \bar c_1^{d}=-\frac{v_u^2}{2(v_u^2+v_d^2)}+\frac12 \bar c_2^{u,d}
\end{eqnarray}
where $\bar c_2^u=1/(1+Z), \bar c_2^d=Z/(1+Z)$ and $c_3'=0$.
The supernovae study of the KSVZ axion has been given before \cite{Chang93}, and now the DFSZ axion can be studied also with the above couplings.

The next important axion coupling is the axion-photon-photon coupling which has a strong model dependence. For several different very light axion models, they are calculated in \cite{Kim98}. Some of these numbers are shown in Table \ref{tab:charges}.

\begin{table}
\begin{tabular}{cr|rcrr}
\hline
\tablehead{1}{r}{b}{KSVZ}
\tablenote{For the unlikely cases of $Q_{\rm em}=\pm\frac23,\pm 1$, we have $ c_{a\gamma\gamma}=0.72, 4.05$, respectively.  For $(m,m)$, we obtain $c_{a\gamma\gamma}=-0.28$.}
  &
  &
  &
  & \tablehead{1}{r}{b}{DFSZ}
  & \\
\hline
$Q_{\rm em}$ & $c_{a\gamma\gamma}$ & $x=\tan\beta$ & ~ same Higgs for ($q^c,e$) masses,  &  $c_{a\gamma\gamma}$\\
\hline
0& $-1.95$ & any $x$, &  $(d^c,e)$    & $0.72$\\
$\pm\frac13$ & $-1.28$ & any $x$, &  $(u^c,e)$ & $-1.28$\\
\hline
\end{tabular}
\caption{$c_{a\gamma\gamma}$ in several field theoretic models. $(m,n)$ in the KSVZ denotes $m$ copies of $Q_{\rm em}=\frac23$ and $n$ copies of $Q_{\rm em}=-\frac13$ heavy quarks with the same PQ charge.}
 \label{tab:charges}
\end{table}

\subsection{Axion mixing in view of hidden sector}

Even if we lowered some $F_a$ from the GUT scale, we must consider the hidden sector also in SUSY models. In this case, the axion mixing must be considered. For the mixing, there is an important cross theorem on the decay constant: {\it Suppose two axions $a_1$ with $F_1$ and $a_2$ with $F_2$ ($F_1 \ll F_2$) couple to two nonabelian gauge groups whose scales have a hierarchy,  $\Lambda_1 \ll \Lambda_2$. Then, diagonalization process chooses the larger condensation scale $\Lambda_2$ chooses the smaller decay constant $F_1$, and the smaller condensation scale  $\Lambda_1$ chooses the larger decay constant $F_2$} \cite{AxionMix}.
So, just obtaining a small decay constant is not enough. The hidden sector may steal the smaller decay constant. It is likely that the QCD axion chooses the larger decay constant \cite{IWKim06}. In this regard, we point out that the MI-axion with anomalous U(1) always has a large decay constant since all fields are charged under this anomalous U(1).

So, phenomenologically successful axion must need an approximate PQ symmetry.
An approximate PQ global symmetry with a discrete symmetry in SUGRA was pointed out long time ago for $Z_9$ in \cite{Lazarides86}. But $Z_9$ is not possible in simple orbifold compactifications of string models. We may need a $Z_3\times Z_3$ orbifold. From  heterotic string,  approximate PQ symmetry are considered in \cite{ChoiKS07}.

\begin{figure}
\caption{Astrophysical and cosmological bounds from axion experiments \cite{KimRMP}.  }\label{SUSYfig04}
\includegraphics[height=.35\textheight]{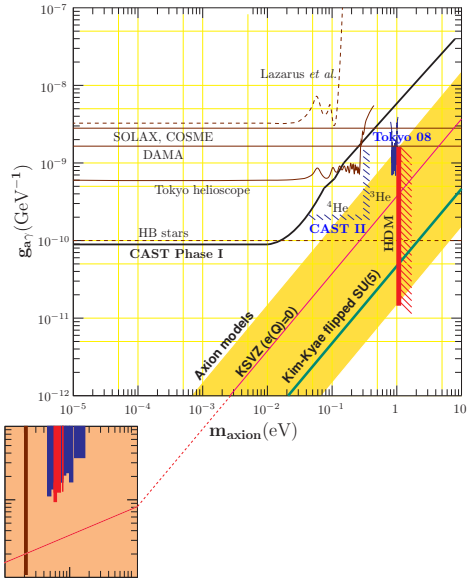}
\end{figure}

In Fig. \ref{SUSYfig04}, we show the current astrophysical and cosmological bounds on the axion decay constant. There exists one calculation \cite{ChoiKS07} from a consistent string model containing a phenomenologically viable MSSM model \cite{KimKyae06}, which is also shown in the figure.

\subsection{White dwarf evolution}

Among the star energy loss mechanisms by the axion emission, the recent white dwarf(WD) study is very interesting. WDs have a very simple structure and easy to study. Their chief aspect is a degenerate electron gas with the temperature $T/\rho^{2/3}< 1.3\times 10^5 K {\rm cm}^2  {\rm gr}^{-2/3}$. For
Sirius B, this number is $3.6\times 10^3$ and it is a WD. The astronomers are able to recover the history of the star formation in our Galaxy by studying the
statistics of WD temperatures. For this, the energy transport mechanism from the
core of a WD is essential. Unlike in Sun, it is transported by neutrinos at high $T$ since most electron are filling the degenerate energy levels.  So, the transport mechanism is very  simple. And the resulting luminosity at the surface
is calculable and reliable.

The later stage of the WD evolution is cristalization from the core. As time goes on, the luminosity drops. In terms of time $t$, the luminosity is $L_{\rm WD}=L_0(1+\frac{t}{\tau_0})^{-7/5}$ where the characteristic time scale $\tau_0$ can be taken as $\tau_0\simeq 2.16\times 10^7$ yrs, for example. A more complete treatment changes this simple behavior little bit \cite{WDloss87}.
The energy loss in the early stage of a WD is through the photon conversion to neutrino pairs in the electron plasma. This calculation of the photon decay was initiated in 1960s, but the accurate number was available after 1972 when
the weak neutral current interaction was taken into account \cite{WDNC72}.

With more data, recently Isern {\it et al.} gives a very impressive figure from the most recent calculation of the above pioneering works, including the early stage and the crystalization period \cite{Isern08}. We translate their number to the axion-electron coupling strength for any axion model \cite{BaeHKKV08} and the resulting axion-electron Yukawa coupling,
\begin{equation}
\left|\frac{m_e\Gamma(e)}{F}\right|\simeq 0.7\times 10^{-13},\quad \frac{m_e\Gamma(e)/N_{\rm DW}}{F_a}\bar ei\gamma_5 e a
\end{equation}
where $F=N_{\rm DW}F_a$ and $\Gamma(e)$ is the PQ charge of electron.

One can think of less exotic processes like the effects of the neutrino transition magnetic moments(NTMM, $\mu$). For the NTMM and the weak neutral currents, one can consider the plasmon decay to neutrinos as \cite{RaffeltBk},
\begin{eqnarray}
&& \Gamma_{\rm NTMM}=\frac{\mu^2}{24\pi}Z_{T,L}\frac{(\omega_{T,L}^2- \omega_{plasmon}^2)^2}{\omega_{T,L}}\\
&& \Gamma_{\rm vector~NC}=\frac{G_F^2C_V^2}{48\pi^2\alpha_{\rm em}^2}Z_{T,L}\frac{(\omega_{T,L}^2- \omega_{plasmon}^2)^3}{\omega_{T,L}}.
\end{eqnarray}
So, the ratio of radiation rates is expressed in terms of $Q$ values (the
convolution of the decay rates and the plasmon distribution
function). Ref. \cite{RaffeltBk} gives $Q_{\rm NTMM}/Q_{\rm SM}=O(1)$ for $\mu\sim 10^{-11}$ times the Bohr magneton. So, the problem here is of the extremely small NTMM ($\ll 10^{-11}\mu_{\rm Bohr}$) in the SM \cite{KimJE78}.

On the other hand, one may introduce new hypothetical light particles: (i) some kind of pseudo-Goldstone bosons considered by Haber \cite{HaberSUSY09}, (ii) a massless or almost massless extra-photon with a kinetic mixing but without the electron coupling to the extra-photon, (iii) a sub-keV milli-charged particle, and (iv) a very light axion.  For Case (i), the pseudo-Goldstone boson coupling to electron is required to be $(3-4)\times 10^{-13}$. For Case (ii), the extra-photon cannot be a candidate for the energy takeout of WDs. For Case (iii), the WD allowed parameter region of Ref. \cite{Davidson00}  is hardly achievable from the red giant bound  \cite{Dienes97}. Case (iv) is commented below.

\begin{figure}
\caption{Positron excess near Earth: (a) scattering of two DMs, $\chi+N$, for the PAMELA $e^+$ \cite{HuhKK08} and (b) a decaying DM $N$ \cite{HuhKimKyae09} for the PAMELA $e^+$ and the {\it Fermi} LAT $e^++e^-$.}\label{SUSY05ab}
\includegraphics[height=.2\textheight]{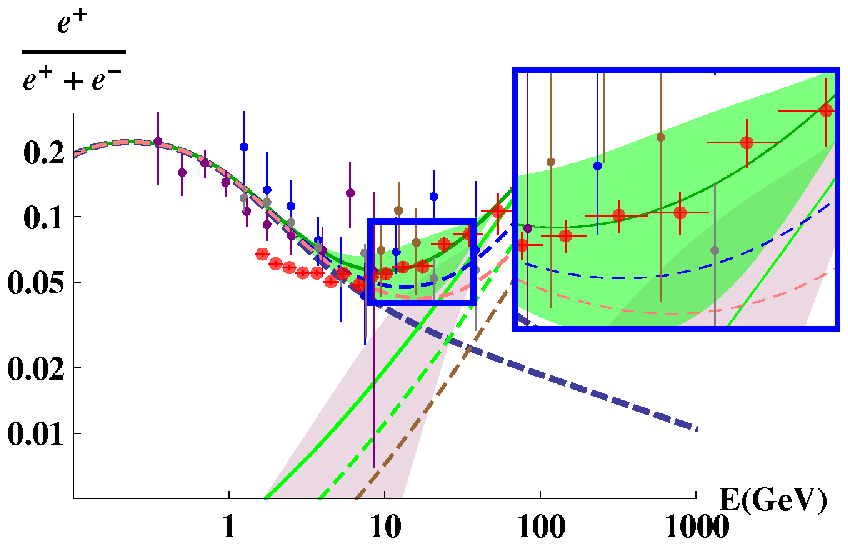}\hskip 0.6cm
\includegraphics[height=.2\textheight]{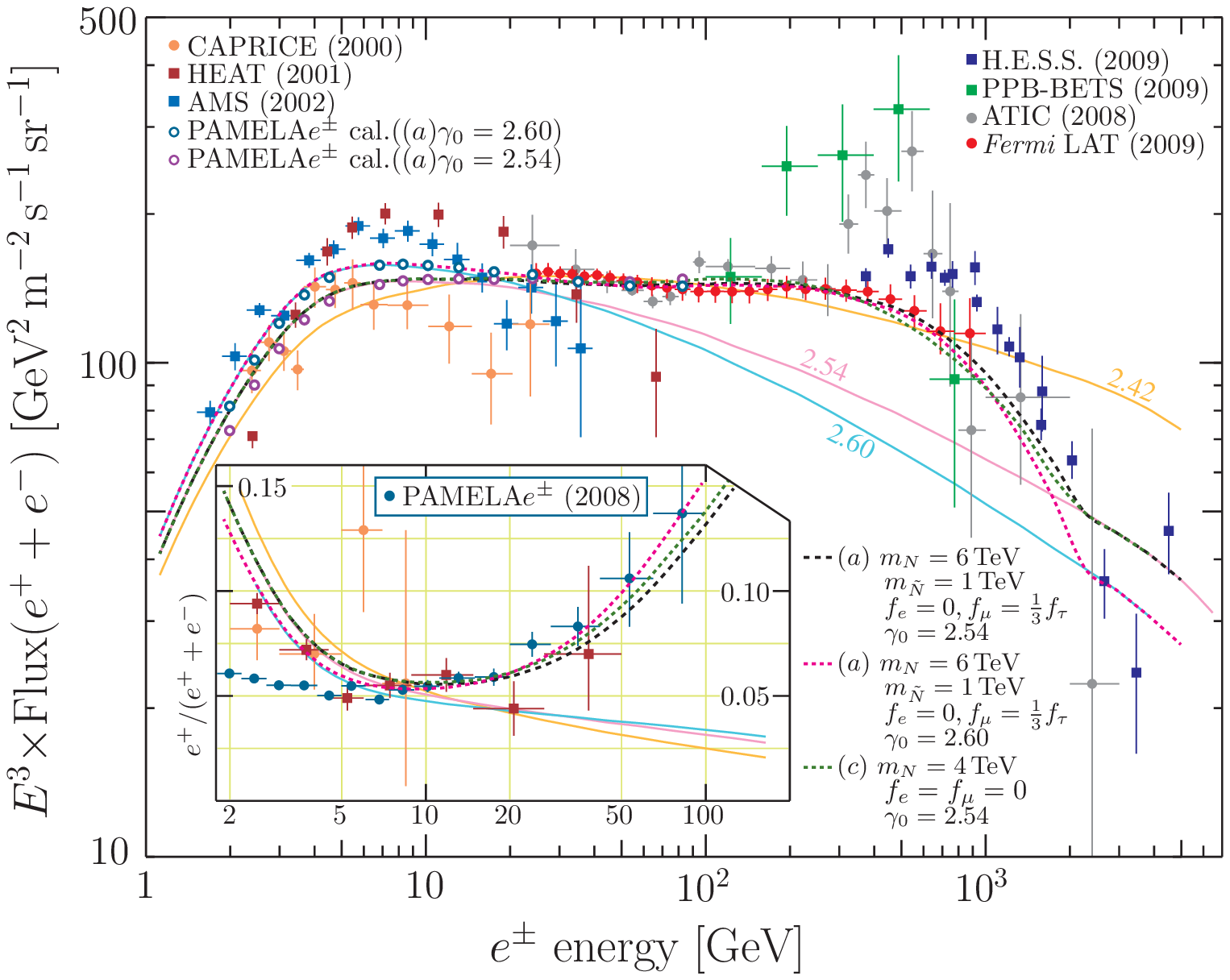}
\hskip -6.5cm
{\small (a) }\hskip 6cm {\small (b)}
\end{figure}

To have a QCD axion at the axion window of $F_a\simeq 10^9 - 10^{12}$ GeV, we need some PQ charge carrying scalar developing VEV(s) at that scale. An enhanced electron coupling compared to the axion
lower bound is possible by,
\begin{enumerate}
  \item[($a$)] Assigning a large PQ charge to $e$. But
      the quark-lepton unification makes this idea not
      very promising, especially in GUTs.
  \item[($b$)] Assign 1 PQ charge to $e$, but let the $N_{\rm DW}$
       be fractional. In this case, only $\frac12$ is possible.
       For the quark sector, effectively e.g. only one chirality
       of one quark carries the PQ charge. But both $e_L$ and $e_R$
       carry the PQ charges.
\end{enumerate}
Case ($b$) has been discussed in  \cite{BaeHKKV08}, where
only $u_R$ is the PQ charge carrying quark. This kind of model is  possible in the flipped SU(5) since $(u, \nu_e, e)^T_L, u_R$ and $e_R$ representations are independent.  In addition, the flipped SU(5) model is good in that it is obtained from string compactification \cite{Antoniadis89,KimKyae06} and also introduces $e_R$ as a GUT singlet, making it possible to interpret the recent leptophilic PAMELA data. This sideway comment is that this flipped SU(5) provides a
two DM scenario for the PAMELA positrons \cite{PAMELAe} through our singlet $e_R$ which is promoted to a heavy charged lepton $E$. Ref. \cite{HuhKK08} was the earliest scattering model with  a leptophilic property while saving neutralino $\chi$ as a DM component. Because the PAMELA data is so intriguing, here we show this idea very briefly from the flipped SU(5). Here, DMs are neutralino $\chi$ and a neutral chiral fermion $N$  with the superpotential $W= fN_R E^c_R e^c_R+ N_R^3.$ In this case, a large enhancement factor is not needed. In the same model without the $N_R^3$ term and raising the mass of $E$ to a GUT scale, the possibility of decaying DM $N$ has been studied also \cite{HuhKimKyae09} in view of the PAMELA  and the {\it Fermi} LAT data \cite{Fermi1}. For a decaying DM $N$, its number density can be calculated in models with a very heavy axino \cite{ChoiKY08}. Since the WD axions and the axino are present in models of  the MSSM extension with the PQ symmetry, these are all in the same framework. In Fig. \ref{SUSY05ab}, we show the allowed parameter regions of these studies.

\subsection{Axions in the universe}

The approximate form of the axion potential is of the cosine function, $-\cos\theta$. The minimum is at $\theta=0$. But the axion potential is almost extremely flat and hence the vacuum stays at any $\theta$ for a long time. It starts to oscillate when the Hubble time $1/H$ is larger than the oscillation
period of the classical axion field $\langle a\rangle$, $3H<m_a$.
This occurs when the temperature is about $0.92$ GeV \cite{BaeFa08}.
In other words, the axion is created at $T\sim F_a$. Since then on the classical
field  $\langle a\rangle$ oscillate. Harmonic oscillator example suggests,
$m_a^2 F_a^2$ is the energy density $\sim m_a\cdot{\rm (number~ density)}$ which behaves like CDM. Ref. \cite{BaeFa08} studies the axion energy density carefully with the axion field evolution equation for a time-varying Lagrangian and the adiabatic condition for the adiabatic invariant quantity, finding an overshoot factor of 1.8.

Thus, if axion is a significant CDM component of the universe \cite{RevEarly}, then it can be detected \cite{Sikivie83}. The experimental efforts for this is reviewed in \cite{KimRMP}. Here, the feeble coupling is compensated by a huge
number of axions, and there is a hope to detect a $10^{-5}$ eV axion. The limits from cosmic and astrophysical axion searches are shown in the previous figure, Fig. \ref{SUSYfig04}. The future ADMX and CARRACK will cover the
interesting region.

\section{Conclusion}

I reviewed axions and the related issues on,
\begin{enumerate}
\item  Solutions of the strong CP problem :
    the $m_u=0$ possibility is ruled out now, the Nelson-Barr type still viable but without a compelling model yet, and the axion solution is most attractive and is not ruled out or may be very difficult to rule out.
\item If axions are discovered by cavity experiments, it will be the case of confirming instanton physics of QCD by experiments, which is most exciting.
\item  Cosmology and astrophysics give bounds on the axion parameters. Maybe, axions are coming out from the WD cooling process. It is the first hint, in the middle of
     the axion window. A specific variant very light axion
     model has been constructed for $N_{DW}=\frac12$.
\item  With SUSY extension, O(GeV) axino can be CDM axino or decaying-to-CDM axino \cite{ChoiKY08}. This kind of axino can produce the needed number of nonthermal neutralinos. In any case, to understand the strong CP with axions in the SUSY framework, the axino must be considered in the CDM discussion, which is presented at this conference   \cite{BaerSUSY09}.
\end{enumerate}

\section{Acknowledgments}
 This work is supported in part by the KRF Grant No. KRF-2005-084-C00001.




\end{document}